\documentclass[twocolumn,english]{revtex4}

\usepackage[latin9]{inputenc}
\usepackage{amsmath}
\usepackage{amssymb}
\usepackage{graphicx}
\usepackage{esint}
\usepackage[usenames, dvipsnames]{color}

\makeatletter
\@ifundefined{textcolor}{}
{%
 \definecolor{BLACK}{gray}{0}
 \definecolor{WHITE}{gray}{1}
 \definecolor{RED}{rgb}{1,0,0}
 \definecolor{GREEN}{rgb}{0,1,0}
 \definecolor{BLUE}{rgb}{0,0,1}
 \definecolor{CYAN}{cmyk}{1,0,0,0}
 \definecolor{MAGENTA}{cmyk}{0,1,0,0}
 \definecolor{YELLOW}{cmyk}{0,0,1,0}
}

\usepackage[english]{babel}
\usepackage{babel}
\usepackage{babel}

\@ifundefined{showcaptionsetup}{}{%
 \PassOptionsToPackage{caption=false}{subfig}}
\usepackage{subfig}
\makeatother

\usepackage{babel}
\begin{document}

\title{Probability of inflation in Loop Quantum Cosmology}

\author{Suzana Bedi\'{c}}
\email{suzana.bedic@icranet.org}

\selectlanguage{english}%

\affiliation{ICRANet, P.le della Repubblica 10, 65100 Pescara, Italy}

\affiliation{ICRA and University of Rome ``Sapienza'', Physics Department, P.le
A. Moro 5, 00185 Rome, Italy}

\author{Gregory Vereshchagin}
\email{veresh@icra.it}

\selectlanguage{english}%

\affiliation{ICRANet, P.le della Repubblica 10, 65100 Pescara, Italy}

\affiliation{ICRA and University of Rome ``Sapienza'', Physics Department, P.le
A. Moro 5, 00185 Rome, Italy}

\affiliation{ICRANet-Minsk, National Academy of Sciences of Belarus, Nezavisimosti
av. 68, 220072 Minsk, Belarus}

\begin{abstract}
We discuss how initial conditions for cosmological evolution can be
defined in Loop Quantum Cosmology with massive scalar field and how
the presence of the bounce influences the probability of inflation
in this theory, compared with General Relativity. The main finding
of the paper is existence of an attractor in the contracting phase of
the universe, which results in special probability distribution at the bounce, quite
independent on the measure of initial conditions in the remote past,
and hence very specific duration of inflationary stage with the number
of e-foldings about $140$. 
\end{abstract}

\maketitle

\section{introduction \label{sec:introduction}}

The concept of inflation has been originally proposed \cite{guth1981inflationary,linde1982anew,albrecht1982cosmology,linde1983chaotic}
as a solution for various cosmological fine tuning, essentially horizon,
homogeneity and flatness problems. An attractive side of inflation
is that it can naturally produce small density perturbations which
permit formation of the observed large scale structure of the universe
out of quantum fluctuations \cite{mukhanov1992theoryof2}. However,
the question about initial conditions for inflation itself remains 
open. Since there is no observational information from the early universe,
one has to consider all possible initial conditions and to draw conclusions
about the probability of inflation. Qualitative analysis of cosmological
solutions in the case of Friedmann-Lemaitre-Robertson-Walker (FLRW)
metric and a massive scalar field $\phi$ performed in \cite{belinsky1985inflationary, 1985PhLB..155..232B}
has indicated the presence of attractor-like trajectories associated
with inflation, showing that inflationary phase in cosmological evolution
is almost generic.

Often pursued idea, as introduced in \cite{gibbons1987anatural},
is to define a Liouville's measure on the phase space of the system,
for instance for massive scalar field in the homogeneous universe,
so called minisuperspace approximation. Despite existing criticism
(see \emph{e.g} \cite{schiffrin2012measure}), measure defined in
this way is widely used, due to lack of information about initial
conditions or physical preference for particular states of the universe.
One of the objections has been the fact that the total measure diverges,
as a result of unbounded volume variable \cite{hawking1988howprobable},
that can be treated in different ways \cite{carroll2010unitary,gibbons2008measure},
possibly leading to different results, for a recent review see \cite{brandenberger2017initial}.
Attractor-like behavior found in \cite{belinsky1985inflationary}
seemed to contradict the Liouville's theorem, which does not allow
for attractor solutions \cite{remmen2013attractor}. This issue has
been clarified in ref. \cite{corichi2014inflationary}, where the
apparent attractor in the phase space $\left(\phi,\dot{\phi}\right)$, where $\dot{\phi}$
is time derivative of $\phi$, is associated with an exponential increase
of the physical volume during inflation. There the measure is defined
on a surface of constant Hubble rate $H$ and then, using Hamiltonian
constraint, is reduced to the form 
\begin{equation}
\varOmega=g\left(H,\phi\right)\mathrm{d}\phi\mathrm{d}v,
\end{equation}
where $v$ is volume of the universe. Probability of some states is
given by the ratio of the integrals of $\varOmega$ over a part of
the phase space containing those states to the same integral taken
over the whole phase space. Divergence in both integrals is avoided
by imposing a cut-off in volume variable. Conclusion drawn from the
analysis is that inflation is generically highly probable for various
potentials and measures.

General relativity (GR) itself does not prescribe how to set up initial conditions, and
it was proposed in \cite{belinsky1985inflationary} that the boundary
of applicability of classical solutions (defined from the equality
of the energy density to Planck's value) can be chosen as the place
where initial conditions should be defined with equal probability
in each point. The discussion on possible choices of the measure is
given in \cite{1987ZhETF..93..784B}, where in particular choices
made in \cite{belinsky1985inflationary} and \cite{gibbons1987anatural}
are compared. The conserved measure $\mathcal{F}$ is defined by the vanishing divergence
\begin{equation}
\partial_i\left(\mathcal{F}v^i\right)=0,\label{measuredef}
\end{equation}
where $v^i$ is Hamiltonian flow vector. The measure of the finite bundle is defined by the integral over the part of the hypersurface $\Delta S$ from which the bundle emanates
\begin{equation}
\mu=\int_{\Delta S}\mathcal{F}v^i ds_i,\label{mumeasuredef}
\end{equation}
where $ds_i$ is the hypersurface element. The conservation of such a measure along the flow and the
fact that it is independent of the choice of the initial hypersurface
are ensured by Eq. (\ref{measuredef}). It is argued that
each particular choice of the measure corresponds to a choice of a
particular exact solution of Eq. (\ref{measuredef}) and there exists
no indication on physical ground how this choice should be made. In
other words, as such the choice of the measure is arbitrary and additional
external arguments should be involved in order to specify it uniquely.
In particular, the choice made in \cite{belinsky1985inflationary}
is based on the argument of inapplicability of classical dynamics
beyond the quantum boundary.

Since the Liouville'e measure diverges for the flat $k=0$ cosmological
models \cite{gibbons2008measure,corichi2014inflationary}, qualitatively
different approach is introduced by Remmen and Carroll \cite{remmen2013attractor}
in context of massive scalar field in flat FLRW universe within GR.
Instead of defining the measure on constant density surface they showed
in which sense $\left(\phi,\dot{\phi}\right)$ space can be regarded as effective
phase space. Considering the measure density $f$ in this effective
phase space they demonstrated that such a measure exists for single
scalar field cosmologies, and that apparent attractor behavior corresponds
to the divergence of the measure $f$, which satisfies the same type of equation as Eq. \eqref{measuredef} for the induced Hamiltonian flow vector $\mathbf{v}$.

Inflation, being very powerful idea in resolving several problems
of the Big Bang cosmology, does not provide a solution for initial
singularity problem \cite{borde2003inflationary}, despite some set
of nonsingular solutions within GR exists, \emph{e.g.}
\cite{page1984afractal}. Loop quantum gravity \cite{thiemann2003lectures,ashtekar2004background,rovelli2004quantum}
is a nonperturbative background independent quantization of GR. Application
of its techniques to homogeneous systems is called loop quantum cosmology
(LQC) \cite{bojowald2008loopquantum}. 
On the genuine quantum level as a sort of general regularity result one can consider a global boundedness of the matter energy density operator in LQC. In terms of trajectories the regularity and in particular the presence of the bounce has been established for several example models (varying in matter content and the topology of Cauchy slices), which include in particular the
model of \cite{ashtekar2006quantum,ashtekar2006quantum2}.
More general results have been obtained on the level of the so called classical effective dynamics. In particular, for the model admitting a massive scalar field with quadratic potential the regularity has been shown in \cite{singh2006nonsingular}. The
closest to general statement in the context of isotropic LQC is the content of
work \cite{2009CQGra..26l5005S}.
Hence in LQC the quantum boundary is naturally replaced by a bounce
on which the energy density of the universe reaches maximum. One may
attempt to define the measure of inflationary solutions at the bounce.
The first treatment of the measure problem in LQC along these lines
is given in \cite{ashtekar2010loopquantum} following the approach
in \cite{corichi2014inflationary}, but with different regularization
of infinities. Allowing all possible initial conditions at the bounce
and constructing dynamical trajectories numerically, it is found in
ref. \cite{ashtekar2010loopquantum} that the probability of inflation
with more than 68 e-foldings (necessary to explain observations) is
almost unity for massive scalar field in flat FLRW universe. Comparison
of their result and explanation of the apparent contradiction with
the corresponding one for GR, obtained in ref. \cite{gibbons2008measure},
is given in ref. \cite{corichi2011measure}. However, setting initial
conditions at the bounce might look an artificial choice.

As the main feature of LQC cosmology with respect to GR is the presence
of the bounce, the question about prebounce cosmological evolution,
when the universe is contracting, arises. Generally speaking, one
can expect that contracting universe is highly inhomogeneous and anisotropic
\cite{1989NYASA.571..249P}. In absence of reliable description of
such contracting phase, following \cite{linsefors2013duration,bolliet_clarifications_2017},
we adopt the same cosmological equations of semiclassical LQC and
study the possibility to set up initial conditions in the remote past
of this cosmological model, namely in the contracting phase prior
to the bounce, characterized by the oscillatory behavior of the scalar
field, and analyze the consequences, in particular for the generality
of inflation.

Our paper is organized as follows. In the next section we introduce
the system and relevant equations, summarizing the effective dynamics
in the LQC perspective. In Sec. \ref{sec:III} we discuss different
choices of the measure on the set of initial conditions and explore
three different choices of the initial probability distributions.
We discuss and interpret obtained results in Sec. \ref{sec:discussion}
which also concludes the paper.

\section{Effective dynamics in LQC}

\label{sec:II}

In LQC the gravitational sector of the phase space is denoted by two
conjugate variables, connection and the triad, which encode curvature
and spatial geometry, respectively (see \emph{e.g.} a review \cite{2011CQGra..28u3001A}).
In this work we consider only flat homogeneous and isotropic spacetime
in which case the dynamical part of the connection is determined by
a single quantity labeled $\mathfrak{c}$ and likewise the triad by
a parameter $p$, related by the Poisson bracket $\left\{ \mathfrak{c},p\right\} =8\pi G\gamma/3$,
where $\gamma$ is the Immirzi parameter whose value $\gamma\thickapprox0.2375$
is set by black hole entropy calculation \cite{thiemann2003lectures,ashtekar2004background,rovelli2004quantum,ashtekar2006quantum,domagala2004blackhole,meissner2004blackhole}.
Relation with the usual GR scale factor $a$ is

\begin{equation}
\mathfrak{c}=\gamma\dot{a},\qquad p=a^{2}.
\end{equation}
 We set $c=1$ and use Planck units; Planck length $l_{\mathrm{{\scriptscriptstyle \textrm{P}l}}}=\sqrt{\hbar G}$
, mass $M_{\mathrm{{\scriptscriptstyle \textrm{P}l}}}=\sqrt{\hbar/G}$
and density $\rho_{\mathrm{{\scriptscriptstyle \textrm{P}l}}}=1/\hbar G^{2}$
.

While the gravitational variables $\mathfrak{c}$ and $p$ are directly
related to the basic canonical pair in loop quantum gravity, the LQC
equations are simpler in the coordinates

\begin{equation}
b=\frac{\mathfrak{c}}{\sqrt{|p|}}=\gamma H,\qquad\mathfrak{v}=\frac{|p|^{3/2}}{2\pi G\hbar\gamma}=\frac{a^{3}}{2\pi G\hbar\gamma},\label{vb_def}
\end{equation}
related by the Poisson bracket $\left\{ b,\mathfrak{v}\right\} =2/\hbar.$
$H=\dot{a}/a$ is the Hubble rate.

Up to a good approximation, the quantum dynamics of LQC can be described
as an effective theory \cite{singh2005semiclassical,taveras2008corrections,banerjee2005discreteness}
generated by an effective Hamiltonian constraint

\begin{equation}
\mathcal{H}_{{\scriptscriptstyle \mathrm{eff}}}=-\frac{3\hbar}{4\gamma\lambda^{2}}\mathfrak{v}\sin^{2}\left(\lambda b\right)+\mathcal{H}_{{\scriptscriptstyle \mathrm{M}}},
\end{equation}
here $\lambda^{2}=4\sqrt{3}\pi\gamma l_{\mathrm{{\scriptscriptstyle \textrm{P}l}}}^{2}$
is the 'area gap', the lowest eigenvalue of the area operator in loop
quantum gravity, and $\mathcal{H}_{{\scriptscriptstyle \mathrm{M}}}$
is the matter Hamiltonian. In our case of a massive scalar field $\phi$
with its canonical momentum $\Pi_{\phi}=\dot{\phi}a^{3}$ and potential
$V\left(\phi\right),$ it is given by

\begin{equation}
\mathcal{H}_{{\scriptscriptstyle \mathrm{M}}}=\frac{1}{2}\frac{\Pi_{\phi}^{2}}{a^{3}}+a^{3}V\left(\phi\right)=a^{3}\rho.
\end{equation}
Energy density and pressure of the scalar field are, as in the classical
case, given by

\begin{equation}
\rho=\frac{1}{2}\dot{\phi}^{2}+V\left(\phi\right),\qquad p_{\phi}=\frac{1}{2}\dot{\phi}^{2}-V\left(\phi\right).
\end{equation}
Using Hamilton's equations for $\dot{\phi}$ and $\dot{\Pi}_{\phi}$
the equation for the scalar field is obtained 
\begin{equation}
\ddot{\phi}+3H\dot{\phi}+V'\left(\phi\right)=0,\label{eq:pfi dot}
\end{equation}
where $V'\left(\phi\right)=\partial V/\partial\phi$. Hamilton's equation
for $\mathfrak{v}$ is 
\begin{equation}
\dot{\mathfrak{v}}=\left\{ \mathfrak{v},\mathcal{H}_{{\scriptscriptstyle \mathrm{eff}}}\right\} =-\frac{2}{\hbar}\frac{\partial\mathcal{H}_{{\scriptscriptstyle \mathrm{eff}}}}{\partial b}=\frac{3}{\gamma\lambda}\mathfrak{v}\sin\left(\lambda b\right)\cos\left(\lambda b\right).
\end{equation}
Using the definition for $\mathfrak{v}$ \eqref{vb_def} we can rewrite
$\dot{\mathfrak{v}}$ as 

\begin{equation}
\dot{a}=\frac{a}{\gamma\lambda}\sin\left(\lambda b\right)\cos\left(\lambda b\right).\label{eq:combining that}
\end{equation}
Now, combining \eqref{eq:combining that} with the vanishing Hamiltonian
constraint, $\mathcal{H}_{{\scriptscriptstyle \mathrm{eff}}}=0$,
we get modified Friedmann equation \cite{singh2006nonsingular,ashtekar2006quantum2}

\begin{equation}
H^{2}=\frac{8\pi G}{3}\rho\left(1-\frac{\rho}{\rho_{c}}\right),\label{eq:H}
\end{equation}
with the critical density 
\begin{equation}
\rho_{c}=\frac{3}{8\pi G\gamma^{2}\lambda^{2}}=\frac{\sqrt{3}}{32\pi^{2}\gamma^{3}}\rho_{\mathrm{{\scriptscriptstyle \textrm{P}l}}}.
\end{equation}
For $\gamma\thickapprox0.2375$ we have $\rho_{c}\thickapprox0.41\rho_{\mathrm{{\scriptscriptstyle \textrm{P}l}}},$
used in numerical calculation.
We will consider scalar field with the mass $m$ and potential

\begin{equation}
V\left(\phi\right)=\frac{m^{2}}{2}\phi^{2}.
\end{equation}

\begin{figure}[ht]
\centering \includegraphics[width=0.4\textwidth]{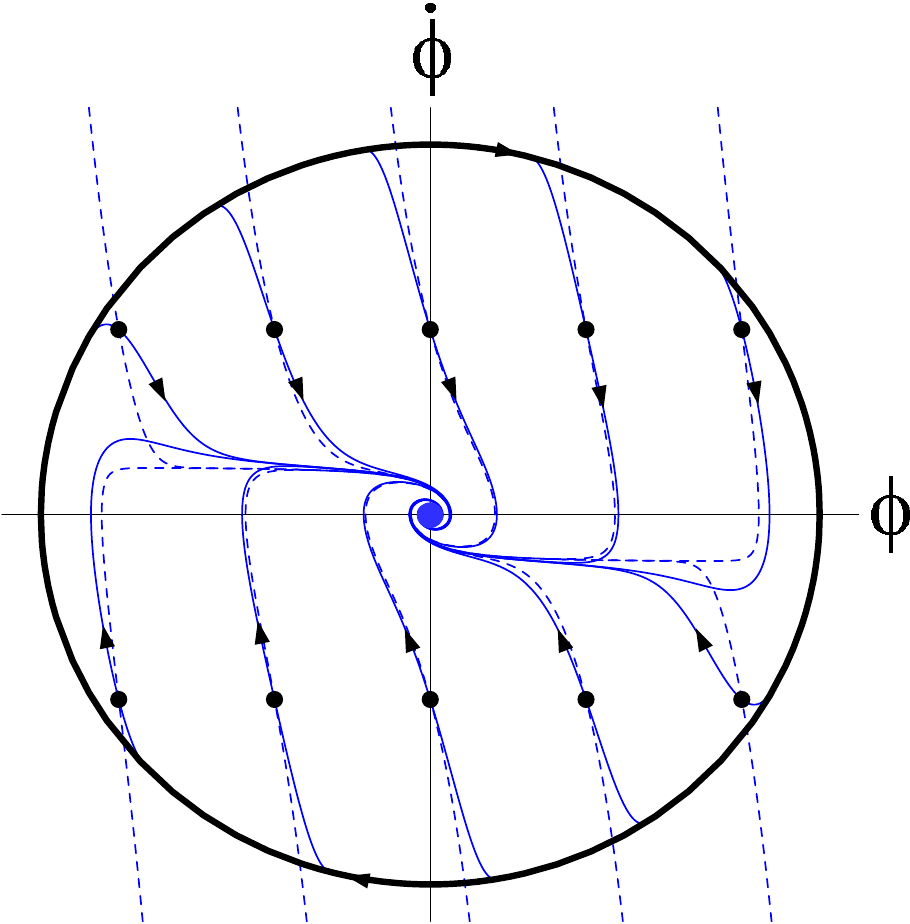} \caption{GR versus LQC inflationary solutions. Reproduced from Ref. \cite{singh2006nonsingular}.}
\label{fig:ph1} 
\end{figure}

The main difference from the GR is realized as an additional term,
quadratic in energy density of the scalar field, in Friedmann equation
(\ref{eq:H}). It comes out that a quantum geometric effect is negligible
for small density, and in the limit $\rho\ll\rho_{c}$ one recovers
ordinary Friedmann equation. In fact, in this limit, where the analysis
of \cite{belinsky1985inflationary} is valid, $H$ can be expressed
from (\ref{eq:H}) and substituted into (\ref{eq:pfi dot}) leaving
only two independent variables: $\phi$ and $\dot{\phi}$. That is
also possible to do in the bouncing cosmologies (with two independent
$\phi-\dot{\phi}$ diagrams for both positive and negative values
of $H$), see \emph{e.g.} \cite{vereshchagin2003flatcosmological}.
However, a quantum geometric effect has a strong effect when density
is comparable to the critical one, see Fig. \ref{fig:ph1}. Both density
and Hubble rate are bounded from above and, going back in time, instead
of running into singularity with diverging physical variables, Hubble
rate vanishes as density approaches the critical value and universe
undergoes regular evolution through the bounce \cite{bojowald2001absence}

\section{Choices of the measure}

\label{sec:III}

As we pointed out in the introduction, LQC incorporates a regular
evolution through the bounce and a description of the prebounce history.
Thus, within LQC it is natural to set up initial conditions in the
remote past \cite{linsefors2013duration}, which in homogeneous and
isotropic case corresponds to the contraction with oscillating scalar
field. As there is no unique way to choose the measure on the initial
surface, in what follows we consider three different choices of measure.

One may argue for the naturalness of the particular definition, as
done in \cite{gibbons1987anatural} for a measure obtained from the
symplectic form of a phase space. This measure diverges for a flat
universe, but the field variables completely specify the system, defining
the effective two-dimensional phase space, and providing a way of
finding on it the unique measure, conserved under the Hamiltonian
flow vector. This is the approach developed in \cite{remmen2013attractor}
and is the base of our first choice of measure.

Another possibility is commonly assumed flat probability distribution.
Particularly, we take the phase of the field on the initial surface
as a natural random parameter, in accordance with the arguments in
\cite{belinsky1985inflationary} and assumed in \cite{linsefors2013duration}
with whom we compare our results. And, finally, the third distribution
we consider is arbitrarily chosen step function in the angle variable.

\subsection{Remmen and Carroll measure}

In this subsection we introduce the notion of an effective phase space
and probability measure defined on it, following Remmen and Carroll
\cite{remmen2013attractor}. The question they discuss is if $\left(\phi,\dot{\phi}\right)$
space could indeed be considered as an effective phase space and how,
if at all, unique conserved measure can be defined on it. We present
here the line of thought and main results, referring the reader to
\cite{remmen2013attractor} for the full derivation and accompanying
discussion.

Being a $2n$-dimensional symplectic manifold, phase space $\Gamma$
has a closed two-form defined on it; 
\begin{equation}
\omega=\sum_{i=1}^{n}\mathrm{d}p_{i}\wedge\mathrm{d}q^{i},
\end{equation}
where $\left(q^{i},p_{i}\right)$ are canonical coordinates and their
conjugate momenta. Liouville's measure is then 
\begin{equation}
\Omega=\frac{\left(-1\right)^{n(n-1)/2}}{n!}\omega^{n}.
\end{equation}
Liouville's theorem states that this measure is conserved along the
Hamiltonian flow vector $X_{\mathcal{H}}$, that can be expressed
by vanishing Lie derivative of $\Omega$, $\mathcal{L}_{X_{\mathcal{H}}}\Omega=0.$
In canonical formulation of GR Hamiltonian is seen as a constraint,
\emph{i.e.} trajectories are confined to a $\left(2n-1\right)$-dimensional
hypersurface $C$ in $\Gamma,$

\begin{equation}
C=\Gamma/\left\{ \mathcal{H}=const.\right\} .
\end{equation}
Evolution of trajectories in $C$ is described by Hamiltonian flow
vector $X_{\mathcal{H}}$ given by 
\begin{equation}
X_{\mathcal{H}}=\frac{\partial\mathcal{H}}{\partial p_{i}}\frac{\partial}{\partial q^{i}}-\frac{\partial\mathcal{H}}{\partial q^{i}}\frac{\partial}{\partial p_{i}}.\label{eq:ham flow vector}
\end{equation}
Space of trajectories then can be defined as $M=C/X_{\mathcal{H}}.$
Unique (under some reasonable conditions) measure on $M$ for FLRW
universe is obtained from the symplectic form $\omega$ by identifying
the $n$th phase space coordinate as time $t$ \cite{gibbons1987anatural},
$\omega=\tilde{\omega}+\mathrm{d}\mathcal{H}\wedge\mathrm{d}t.$ The
corresponding measure is 
\begin{equation}
\Theta=\frac{\left(-1\right)^{\left(n-1\right)(n-2)/2}}{\left(n-1\right)!}\tilde{\omega}^{n-1}.
\end{equation}

We can think of $\left(\phi,\dot{\phi}\right)$ space as an effective
phase space if it captures the entire dynamics of the system. Eliminating
$a$ and $\dot{a}$ from the dynamics, possible only in flat FLRW
model, and expressing $H$ as a function of $\phi$ and $\dot{\phi}$
by use of Friedmann equation, reduces Hamiltonian flow vector \eqref{eq:ham flow vector}
to only two components, $\phi$ and $\dot{\phi}$. It means that if
we consider the map $\chi:C\rightarrow K\cong\left(\phi,\dot{\phi}\right)$
defined by

\begin{equation}
\chi\left(a,\dot{a},\phi,\dot{\phi}\right)=\left(\phi,\dot{\phi}\right)
\end{equation}
it represents a vector field invariant with respect to the Hamiltonian
flow vector $X_{\mathcal{H}}$. Roughly speaking, we have unique (no
intersecting trajectories) induced Hamiltonian vector field on $\left(\phi,\dot{\phi}\right)$, reflecting the behavior of the Hamiltonian vector field of the full
phase space. All points in $3-$dimensional constrained surface $C$
with the same $\left(\phi,\dot{\phi},\,\textrm{sign}(H)\right)$ values
are mapped into one point in $\left(\phi,\dot{\phi}\right)$ space.
In GR the effective phase space is the entire $\phi-\dot{\phi}$
plane. In LQC the phase space is composed of two ellipsoidal sheets, one
for each $\textrm{sign}(H)$, glued together at their boundaries,
corresponding to $H=0$. On both sheets the phase space trajectories do not intersect. So the effective dynamics of LQC can be considered as a Hamiltonian flow on the reduced phase space of nontrivial topology.

We denote induced vector field as $\mathbf{v}$ and define $x=\phi,\,y=\dot{\phi}$.
The measure on $\left(\phi,\dot{\phi}\right)$ is a two-form $\boldsymbol{\mathbf{\sigma}}$
that can be written as 
\begin{equation}
\boldsymbol{\mathbf{\sigma}}=f\left(x,y\right)\mathrm{d}x\wedge\mathrm{d}y
\end{equation}
for some function $f$. Conservation of the measure along $\mathbf{v}$
is provided by $\mathcal{L}_{\mathbf{v}}\boldsymbol{\mathbf{\sigma}}=0,$
which can be expressed as
\begin{equation}
\nabla\left(f\mathbf{v}\right)=0.\label{eq:div(fv)}
\end{equation}
It was shown \cite{1987ZhETF..93..784B,remmen2013attractor} that
if there exist a function $f$ satisfying \eqref{eq:div(fv)}, it
defines the measure on effective phase space uniquely, \emph{i.e.
} 
\begin{equation}
\mathrm{d}\Pi_{\phi}\wedge\mathrm{d}\phi=f\mathrm{d}\dot{\phi}\wedge\mathrm{d}\phi,\label{measureRC}
\end{equation}
where $\Pi_{\phi}$ is momentum conjugate to $\phi$ in $\left(\phi,\dot{\phi}\right)$
space. Equation (\ref{measureRC}) shows that the natural Liouville
measure is equivalent to the measure $f\mathrm{d}\dot{\phi}\wedge\mathrm{d}\phi$
introduced by Remmen and Carroll \cite{remmen2013attractor}.

The vector field $\mathbf{v}$ is given by 
\begin{equation}
\mathbf{v}=\dot{\phi}\boldsymbol{\mathbf{\hat{\phi}}}+\ddot{\phi}\boldsymbol{\hat{\dot{\phi}}},
\end{equation}
where $\ddot{\phi}$ is obtained from \eqref{eq:pfi dot} and $H$
from \eqref{eq:H}. If we define the coordinates 
\begin{equation}
x=\phi=\frac{r}{m}\cos\theta,\qquad y=\dot{\phi}=r\sin\theta,
\end{equation}
we have 
\begin{equation}
\rho=\frac{r^{2}}{2},\qquad H^{2}=\frac{4\pi Gr^{2}}{3}\left(1-\frac{r^{2}}{2\rho_{c}}\right),
\label{eq:Hr}
\end{equation}
and vector field $\mathbf{v}$ in polar coordinates:

\begin{equation}
\mathbf{v}=-3rH\sin^{2}\theta\mathbf{\hat{r}}-r\left(m+3H\sin\theta\cos\theta\right)\boldsymbol{\hat{\theta}},
\end{equation}
where we have used standard transformation 
\begin{equation}
\mathbf{\hat{x}}=m\cos\theta\mathbf{\hat{r}}-m\sin\theta\boldsymbol{\hat{\theta}},\quad\;\mathbf{\hat{y}}=\sin\theta\mathbf{\hat{r}}+\cos\theta\boldsymbol{\hat{\theta}}.
\end{equation}
Constraint \eqref{eq:div(fv)} on the measure density $f\left(r,\theta\right)$
then becomes partial differential equation

\begin{equation}
\frac{1}{r}\partial_{r}\left(Hr^{2}f\right)\sin^{2}\theta+\frac{m}{3}\partial_{\theta}f+H\partial_{\theta}\left(f\sin\theta\cos\theta\right)=0.\label{eq:F skra=00003D00003D00003D00003D00003D000107ena}
\end{equation}

Inserting eq. (\ref{eq:Hr}) into eq. (\ref{eq:F skra=00003D00003D00003D00003D00003D000107ena}) one can derive the following result (the corresponding result within the general relativity is obtained in \cite{remmen2013attractor}),

\begin{equation}
f=A\left[\frac{1}{r^3\sqrt{1-\frac{r^2}{2 \rho_c}}}+\frac{2\sqrt{3\pi G}\sin\theta\cos\theta}{mr^{2}}\right],\;A\epsilon\mathbb{R},\label{rho0measure-1}
\end{equation}
with the plus sign coming from the negative $H.$

A measure on the space of trajectories (as opposed to the effective
phase space) can be constructed from the effective phase space measure
on any surface transverse to those trajectories, by demanding that
the physical result be independent of the chosen transverse curve
\cite{remmen2014howmany}. It can be any transverse slicing that evolves
monotonically in time, so we will take $r\,\left(\propto\sqrt{\rho}\right)=const$
surface and parametrize it by the angular coordinate $\theta$. For
a bundle of trajectories centered at the angle $\theta_{1}$ and spanned
by $\textrm{d}\theta_{1}$ on initial surface at the radius $r,$
we can write its probability measure as $P\left(\theta_{1}\right)\textrm{d}\theta_{1}$
and let it evolve to another surface $r=r_{2}.$ Condition for the
measure to be conserved is then 
\begin{equation}
P\left(\theta_{1}\right)|_{r_{1}}\textrm{d}\theta_{1}=P\left(\theta_{2}\right)|_{r_{2}}\textrm{d}\theta_{2}.\label{eq:Pdth}
\end{equation}
Now, for a region $\textrm{d}\theta_{1}\textrm{d}r_{1}$ that evolves
to $\textrm{d}\theta_{2}\textrm{d}r_{2}$, we have Liouville's theorem
for the effective phase space 
\begin{equation}
f\left(r_{1},\theta_{1}\right)\textrm{d}\theta_{1}\textrm{d}r_{1}=f\left(r_{2},\theta_{2}\right)\textrm{d}\theta_{2}\textrm{d}r_{2},\label{eq:fdrdth}
\end{equation}
\emph{i.e.} $f$ satisfies \eqref{eq:div(fv)}. Comparing
\eqref{eq:Pdth} and \eqref{eq:fdrdth} we conclude that the probability
distribution on the space of trajectories, $r=const$ surface, is
$P\left(\theta\right)|_{r}\propto f\textrm{\ensuremath{\left(r,\theta\right)}d}r$.
We could have first divide \eqref{eq:fdrdth} by $\textrm{d}t$ (which
we can do since $t$ evolves uniformly for all trajectories) thus
obtaining

\begin{equation}
P\left(\theta\right)|_{r}\propto f\left(r,\theta\right)|\dot{r}|,
\end{equation}
with the proportionality constant given by the normalization $\intop_{0}^{2\pi}P\left(\theta\right)|_{r}\textrm{d}\theta$=1.
For \eqref{rho0measure-1} and

\begin{equation}
|\dot{r}|=\left|-\frac{3H}{r}\dot{\phi}^{2}\right|\simeq2\sqrt{3\pi G}r^{2}\sin^{2}\theta
\end{equation}
we finally obtain 
\begin{equation}
P\left(\theta\right)|_{r}=\frac{\sin^{2}\theta}{\pi}+\sqrt{\frac{3G}{\pi}}\frac{2r\,\sin^{3}\theta\cos\theta}{m}.\label{cosprob}
\end{equation}
This probability distribution of initial conditions is used below
to infer the generality of inflation.

\subsection{Linsefors and Barrau measure}

Closely following Linsefors and Barrau \cite{linsefors2013duration},
we define again $x=\phi,\,y=\dot{\phi}$ so that $\rho=\frac{1}{2}\left(m^{2}x^{2}+y^{2}\right)$
and equations of motion for $x,$$y$ and $\rho$ are

\begin{equation}
\dot{x}=y,\qquad\dot{y}=-3Hy-m^{2}x,\qquad\dot{\rho}=-3Hy^{2}.\label{mainEq}
\end{equation}
Given the conditions for the prebounce oscillations, 
\begin{equation}
\rho\ll\rho_{c},\qquad|H|\ll m;\qquad H\approx-\sqrt{\frac{8\pi G\rho}{3}},
\end{equation}
we can approximate $x$ and $y$ by 
\begin{equation}
x=\frac{\sqrt{\rho}}{m}\sin\left(mt+\delta\right),\qquad y=\sqrt{\rho}\cos\left(mt+\delta\right),\label{xysol}
\end{equation}
which gives 
\begin{equation}
\dot{\rho}=\sqrt{24\pi G}\rho^{3/2}\cos^{2}\left(mt+\delta\right)
\end{equation}
with the solution 
\begin{eqnarray}
\rho=\rho_{0}\left[1-\sqrt{\frac{3\pi G\rho_{0}}{2}}\left(t+\frac{1}{2m}\sin\left(2mt+2\delta\right)\right.\right.\nonumber \\
\left.\left.-\frac{1}{2m}\sin\left(2\delta\right)\right)\right]^{-2}.\label{eq:densityLB}
\end{eqnarray}

\begin{figure}
\subfloat[\label{fig:-Asymptotic-solution}Only the term $t^{-2}$ is kept.
Blue circle shows the boundary $\rho_{0}$ where initial conditions
are set.]{\includegraphics[scale=0.12]{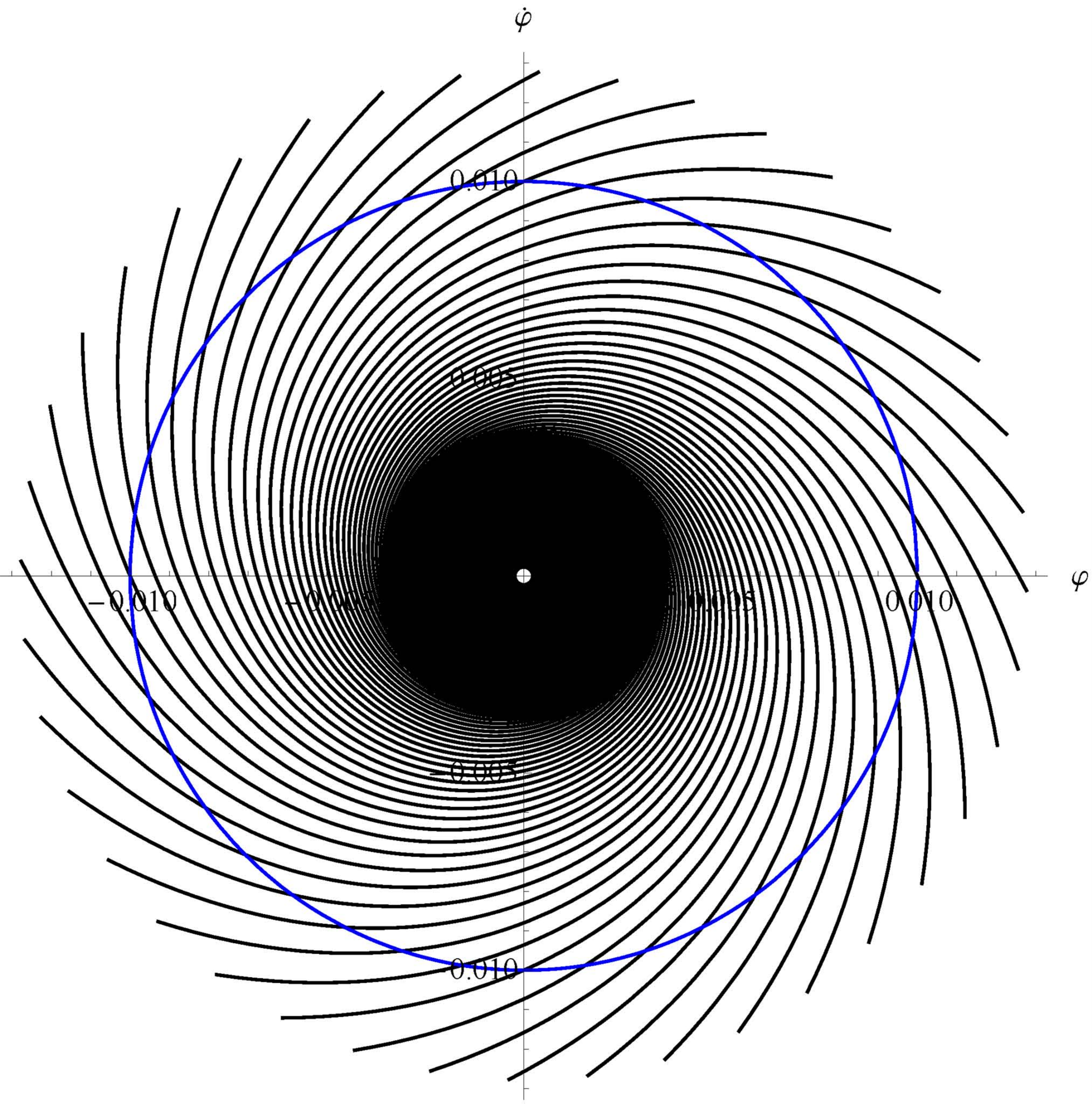}

}

\subfloat[\label{fig:Full-asymptotic-solution}Full asymptotic solution of the
same equations corresponding to Eq. (\ref{eq:densityLB}). Red curves
denote repulsive separatrices.]{\includegraphics[width=0.3\textwidth]{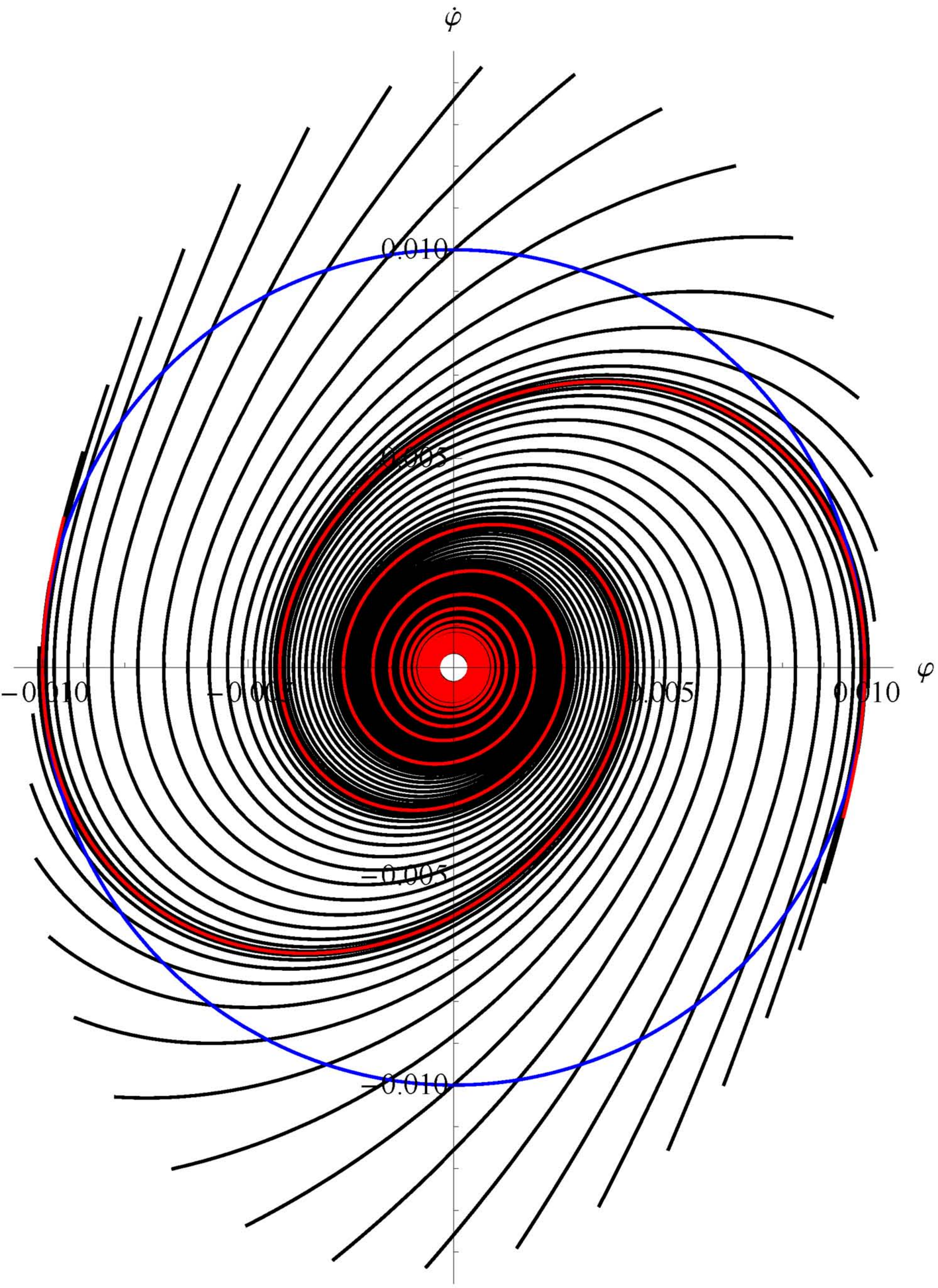} 

}

\caption{\label{fig:densityab}Asymptotic solution of cosmological equations
of LQC with massive scalar field in the limit $\rho\rightarrow0$
corresponding to Eqs. (\ref{xysol}) and (\ref{eq:densityLB}). The
origin is denoted by white circle.}
\end{figure}

Note the presence of an additional term with respect to the corresponding
Eq. $\left(8\right)$ in \cite{linsefors2013duration}. In Fig. \ref{fig:densityab}
we present phase space trajectories on the $\phi-\dot{\phi}$ plane
near the origin, where Eq. (\ref{eq:densityLB}) is valid. In the
limit $t\rightarrow-\infty$ one keeps just the term proportional to
$t^{-2}$ in \eqref{eq:densityLB} (see \emph{e.g. }\cite{belinsky1985inflationary})
corresponding to the solutions shown in Fig. \ref{fig:-Asymptotic-solution},
where the blue circle represents the constant density surface. Full
expression for the density, on the other hand gives rise to a more
complex structure shown in Fig. \ref{fig:Full-asymptotic-solution}.

The assumption on the measure in \cite{linsefors2013duration} is
that the probability of initial conditions does not depend on $\delta$,
namely 
\begin{equation}
P(\delta)=const.\label{eqprob}
\end{equation}
For comparison with the previous subsection, notice $\delta=\pi/2-\theta|_{t=0}$.

Comparing Fig. \ref{fig:densityab} a) and b) one can understand the
emergence of the separatrix at the contraction phase. This separatrix
is repulsive, in contrast to the attractive inflationary separatrix
at the expansion phase. It breaks down the symmetry of solution shown
in Fig. \ref{fig:-Asymptotic-solution} and introduce an additional
dependence of the probability of solutions on the phase $\delta$
and on the mass $m$, see Eq. (\ref{eq:densityLB}).

After the oscillatory phase ends most solutions do not follow this
separatrix and this gives origin to the tendency of the trajectories
to end up with high probability at the same point at the bounce. Although
this has been observed in \cite{linsefors2013duration}, the explanation
was lacking.

\subsection{Narrow distribution}

As a third choice of the measure we adopt for the probability distribution
two short intervals of widths equal to $\pi/20$, with constant probability
within. This choice of the measure is not motivated by any physical
arguments, it just represents a narrow distribution in angle variable
and we intend to compare the probability distribution at the bounce
with other choices of the measure discussed above.

\subsection{Results}

\begin{figure}
\subfloat[\label{fig:Mean}Dependence of the peak of probability density distribution
on scalar field mass.]{\includegraphics[width=0.48\textwidth]{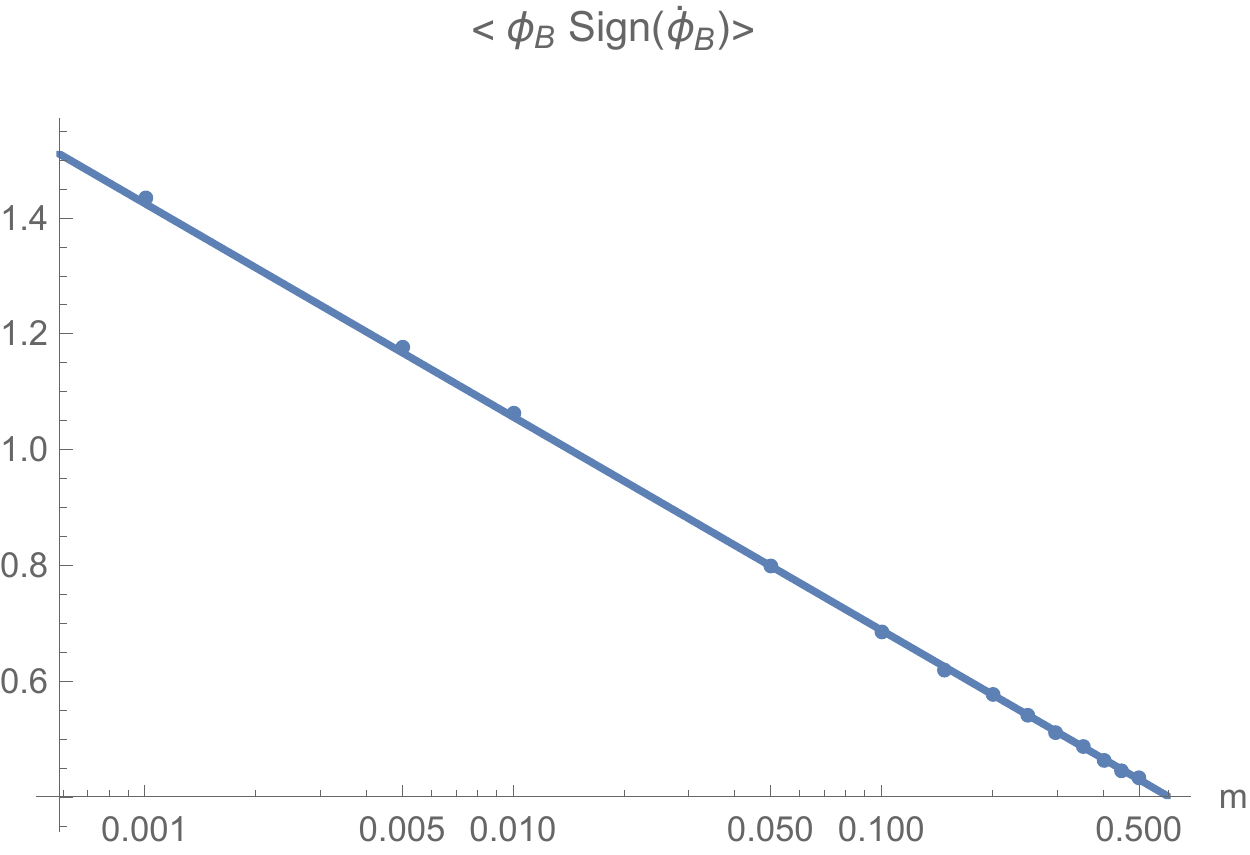}

}

\subfloat[\label{fig:standard deviation}Dependence of the standard deviation
in the probability density distribution on scalar field mass.]{\includegraphics[width=0.48\textwidth]{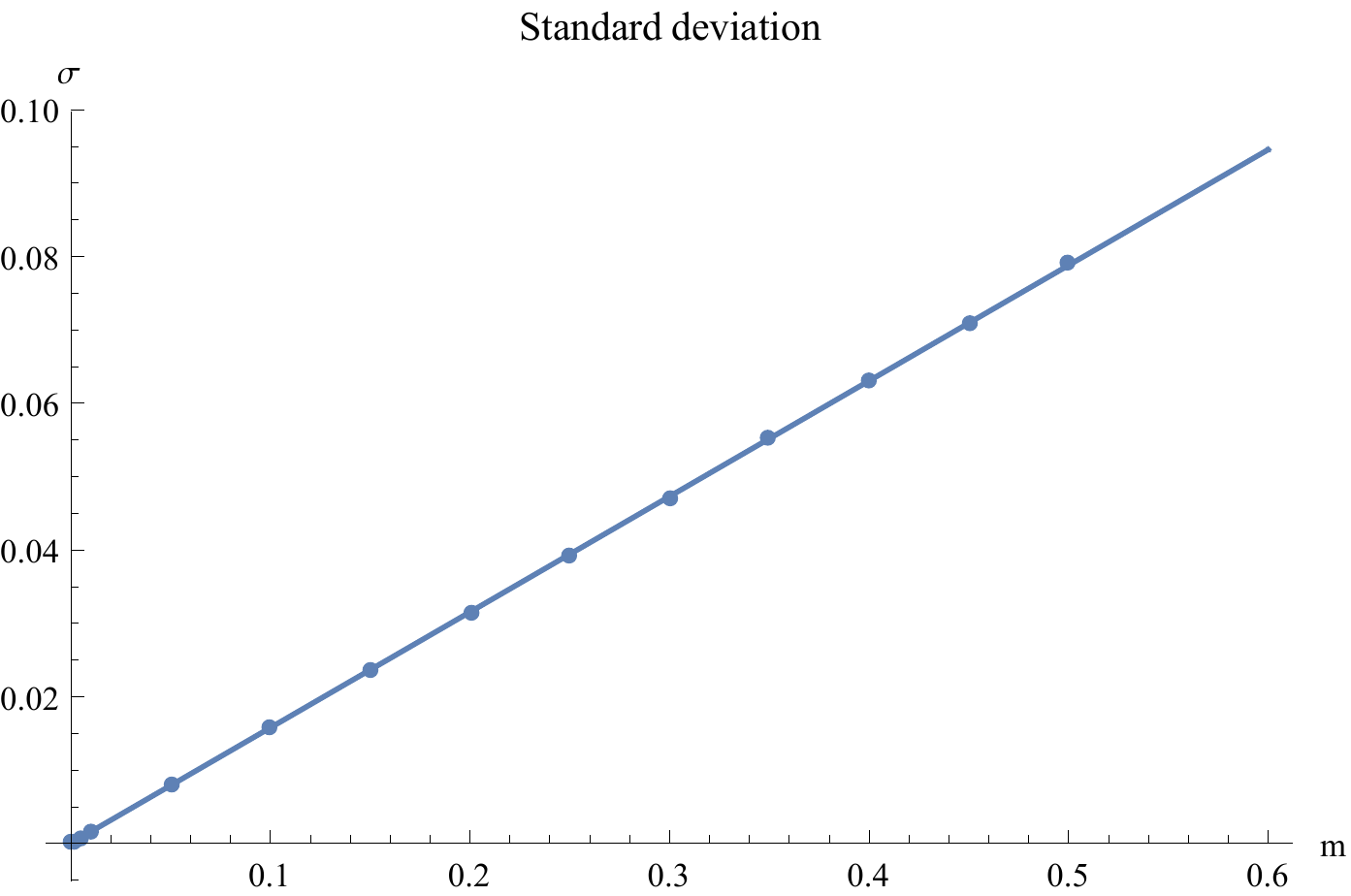} 

}

\caption{Mass is given in Planck units.}\label{fig:MeanStan}
\end{figure}

\begin{figure}
\subfloat[\label{fig:3 P(delta)}Probability distributions for initial conditions
in the remote past given by: Eq. (\ref{eqprob}) - orange, Eq. (\ref{cosprob})
- red, narrow distribution - blue. Vertical lines denote the values of $\delta$ variable used in the calculations.]{\includegraphics[width=0.9\columnwidth]{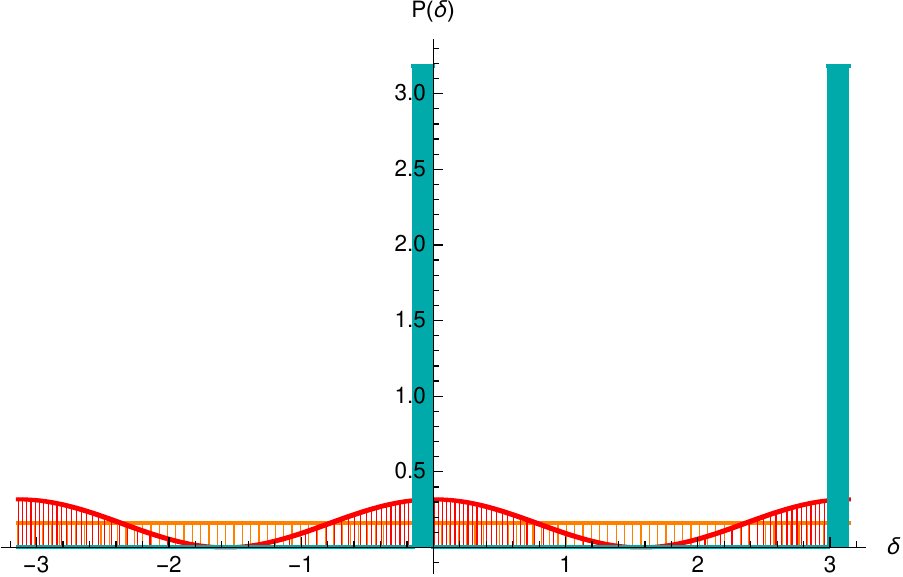}
}

\subfloat[\label{fig:xBounce PDF}Probability distribution of the rescaled field
values at the bounce.]{\includegraphics[width=0.9\columnwidth]{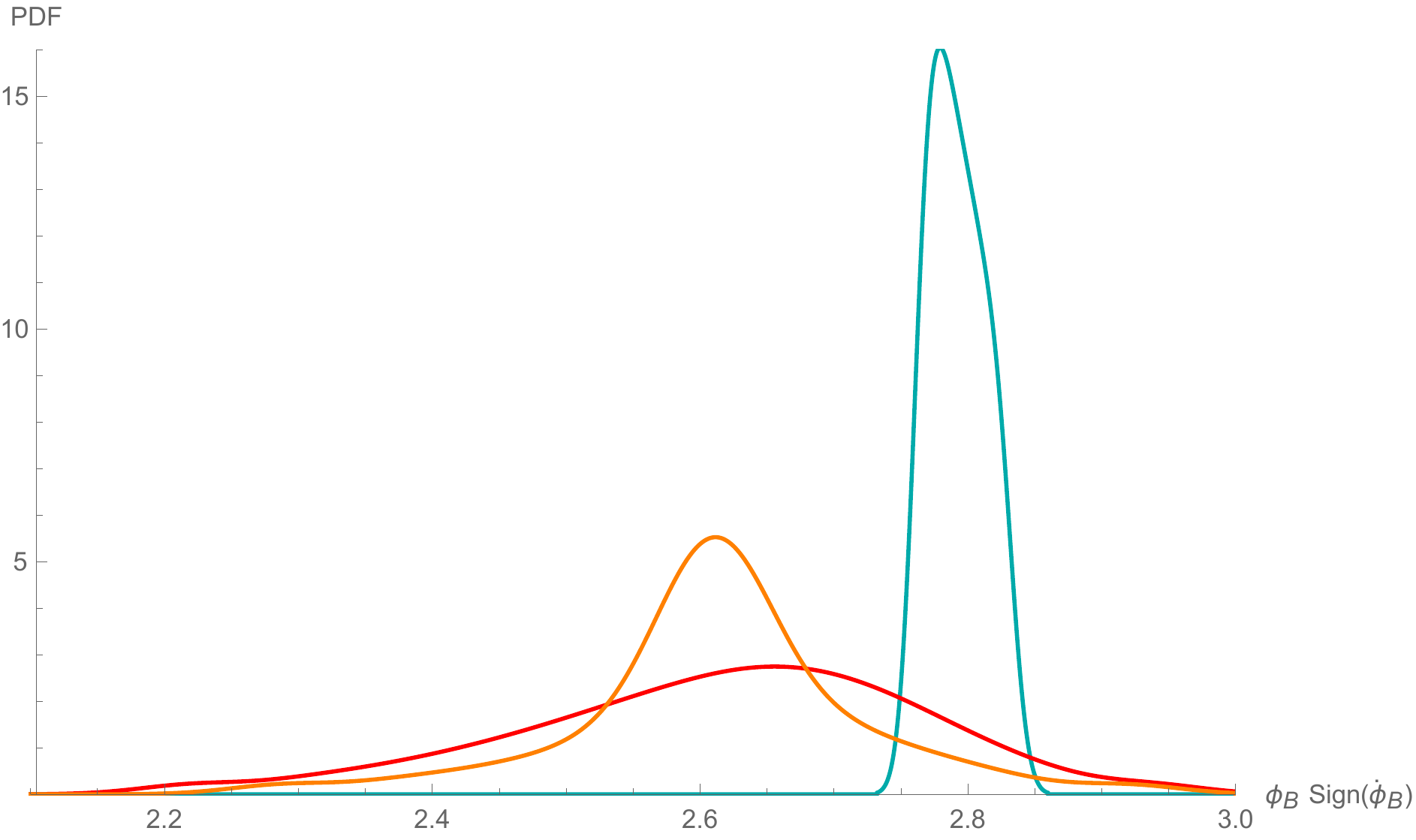} 

}

\caption{Initial probability distributions and probability distributions at
the bounce for all three cases considered.}
\label{probdistr} 
\end{figure}

The set of initial probability density distributions is shown in Fig.
\ref{fig:3 P(delta)}. For the probability distribution \eqref{eqprob}
we examine the dependence of the value of the field at the bounce
on the mass, by choosing initial conditions at some $r_{0}\ll m$,
evolving them up to the bounce and computing the mean $\left\langle ...\right\rangle $
and the standard deviation $\sigma_{B}$ of the quantity $\phi_{B}\,\textrm{sign}\left(\dot{\phi}_{B}\right)$.
Solutions at the bounce can be parametrized by $\phi_{B}$ and $\textrm{sign}\left(\dot{\phi}_{B}\right)$
but, as done in \cite{linsefors2013duration} we project them to the
physically relevant parameters. Repeating the calculation for different
masses while keeping the ratio $r_{0}/m=0.005$, we obtain the result
shown in Fig. \ref{fig:MeanStan}, which can be reasonably fit with
the formulas 
\begin{eqnarray}
\left\langle m\phi_{B}\,\textrm{sign}\left(\dot{\phi}_{B}\right)\right\rangle  & = & m\left(0.32+0.16\,Log\left(\frac{1}{m}\right)\right),\:\;\\
\sigma_{B} & = & 0.16m.
\end{eqnarray}
These results imply that for $m\ll M_{\mathrm{{\scriptscriptstyle \textrm{P}l}}}$ for most trajectories the bounce
occurs with $\sigma_{B}$ and $\phi_{B}\ll M_{\mathrm{{\scriptscriptstyle \textrm{P}l}}}$.

We repeat the process of evolving trajectories and computing the mean
of the field value at the bounce for all three choices of measure
with the fixed mass $m=1.21\cdot10^{-6}M_{\mathrm{{\scriptscriptstyle \textrm{P}l}}}$,
chosen for comparison with \cite{linsefors2013duration}. The obtained
probability density distributions of the variable $\left\langle m\,\phi_{B}\,\textrm{sign}\left(\dot{\phi}_{B}\right)\right\rangle $
at the bounce are shown in Fig. \ref{fig:xBounce PDF}. For initial
conditions with the probability \eqref{eqprob} we find good agreement
with the result of \cite{linsefors2013duration} where the distribution
is found to be sharply peaked around $3.55\cdot10^{-6} M_{\mathrm{{\scriptscriptstyle \textrm{P}l}}}^2.$

Surprisingly, other very different probability distributions also
result in a similar distributions at the bounce, despite so different
choices of the measure in the remote past. The field variable $\left\langle \phi_{B}\,\textrm{sign}\left(\dot{\phi}_{B}\right)\right\rangle $
is narrowly distributed in the region $\dot{\phi}\ll M_{\mathrm{{\scriptscriptstyle \textrm{P}l}}}^2$
at the bounce for all cases considered. Thus, our results show that
there exists a kind of attractor also in the contracting branch, and
the field values on the bounce are virtually independent of the initial
probability distribution, as long at it is smooth.

The number of e-folds during slow-roll inflation can be estimated
following \cite{linsefors2013duration} as 
\begin{equation}
N= 2 \pi \textup{G}  \phi_{max} ^{2},\label{Nest}
\end{equation}
so knowing the maximum value of the scalar field reached after the
bounce we can estimate the value of $N$. Since $\phi_{max}\approx2\phi_{B}$
we find $N\approx170$.

We calculate numerically the number of e-foldings corresponding to
the mean field value at the bounce for the three cases, using 
\begin{equation}
N=\int_{t_{in}}^{t_{end}}H\mathrm{d}t,
\end{equation}
where $t_{in}$ and $t_{end}$ are the initial and ending time of
the inflation, respectively. As the start of the inflation we take
the time when field takes the maximum value after the bounce, and
for the end we take the $|H|=m$ condition. Particularly, we do not
include super-inflation phase into the number of e-folds calculation.
We get $N\approx143$ for the first two, and $N\approx153$ for the
third initial distribution.

Therefore, it is the main result of this paper that the probability
distribution at the bounce strongly depends on mass of the scalar
field, but weakly depends on initial probability distribution in the
remote past.

\section{Discussion and conclusions\label{sec:discussion}}

\begin{figure}
\centering \includegraphics[scale=0.6]{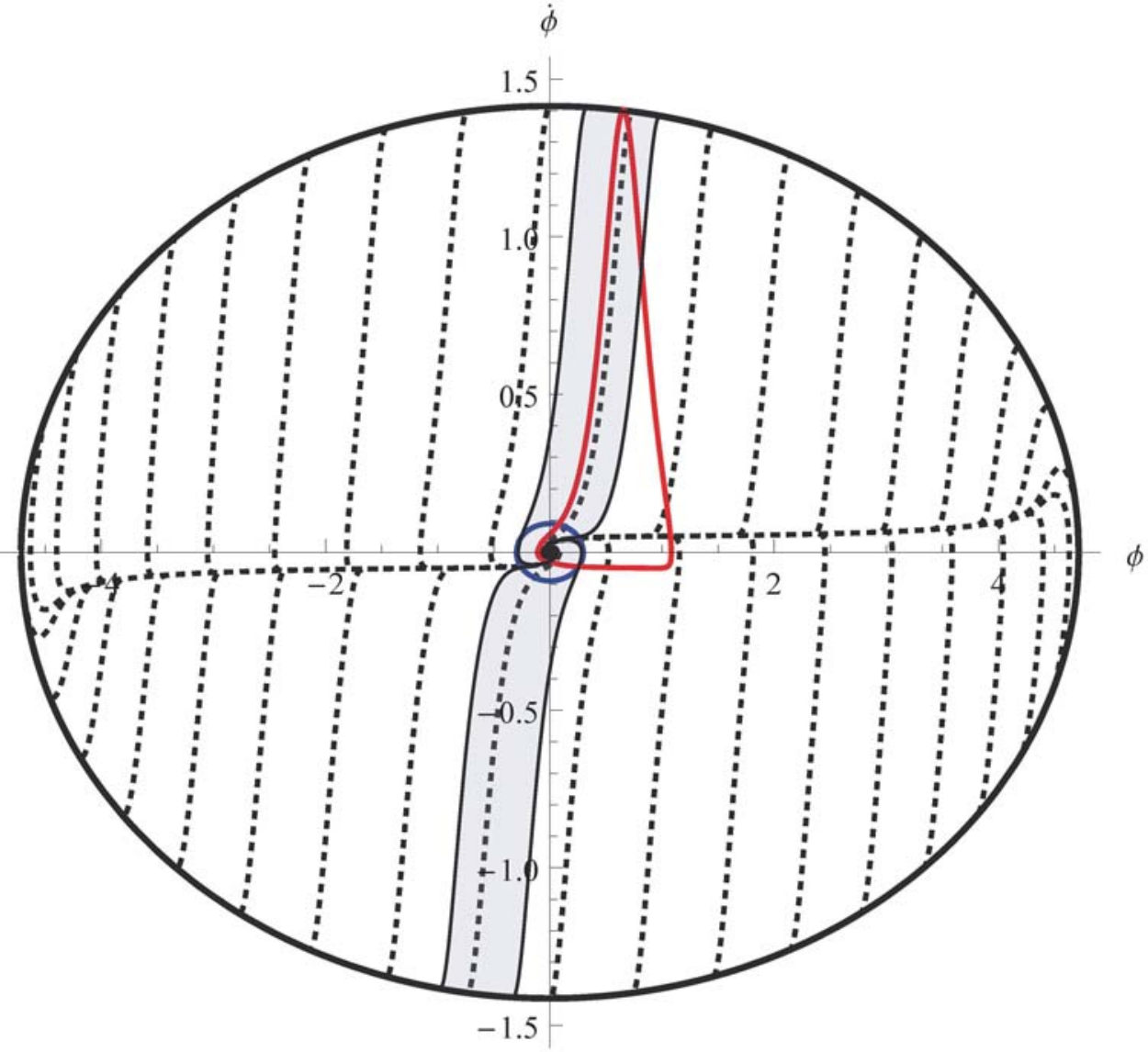} \caption{\label{fig:most probable sol}Solutions of the cosmological equations
of LQC in the $\phi-\dot{\phi}$ diagram. All solutions start in the
distant past in the origin, pass through oscillatory phase with increasing
amplitude of oscillations and end up at the bounce shown by the thick
external circle. Blue internal circle (obtained setting $H=m$) represents
the region where most solutions deviate from the repulsive separatrix
(located along the horizontal axis, at which the Hubble parameter
is nearly constant). This region shrinks with decreasing mass (the
value of the mass in this figure is $m=0.1 M_{\mathrm{{\scriptscriptstyle \textrm{P}l}}}$ selected for better clarity).
The red curve shows the complete most probable solution originating
from the remote past (including both contraction and expansion phases).
This solution has no exponential contraction phase, but it has a successful
inflationary phase.} 
\label{fullsol} 
\end{figure}

The main question arising from the results reported above is why most
solutions originating from the oscillatory contracting phase end up
at the bounce having very restricted values of the scalar field? This
can be understood as a consequence of the presence of repulsive separatrix
at the contraction phase, as well as small value of the mass of the
scalar field, compared with the Planck mass, see Fig. \ref{fig:most probable sol}.

Due to repulsive nature of the separatrix, most solutions, starting at
the origin, do not follow along it, and deviate from it as early as
possible. The separatrix appears at small enough densities, see Fig.
\ref{fig:Full-asymptotic-solution} above. However, all solutions
are located between the pair of repulsive separatrices as long as
the solution is oscillating near the origin $\rho=0$. This behavior
breaks down when $H\sim m$, see Fig. \ref{fullsol}. Then most solutions
cannot propagate to large values of $\phi$ at the contraction phase
and leave the region near the origin in the vertical direction (see
shaded region in Fig. \ref{fig:most probable sol}). This picture
corresponds exactly to Fig. \ref{fig:xBounce PDF}.

Therefore, indeed as pointed out in \cite{linsefors2013duration},
there is a preferred set of cosmological solutions in LQC, which have
no exponential contraction phase, but possess a successful inflationary
phase. This is a direct consequence of the prebounce evolution in
LQC, specifically existence of the repulsive separatrix in contracting
phase. In other words, one may say that there is an attractor behavior
in LQC, which not only ensures the successful inflation, but also
determines the prebounce evolution in LQC. This attractor is shown
in Fig. \ref{fig:most probable sol} by the red curve.

Contrary to the assumptions made in previous
literature,\emph{e.g.} in \cite{ashtekar2010loopquantum,corichi2011measure,corichi2014inflationary},
one consequence of our analysis is a well defined probability distribution
for the scalar field at the bounce. This probability distribution is strongly peaked at some particular scalar field values, see Fig. \ref{probdistr}b. This result hence impacts on
the prediction of the duration of inflation, and it can be subject
to precision tests, such as CMB anisotropy measurements.

The analysis of this paper was based on the simplest quadratic effective
potential for the scalar field. Clearly, many inflationary potentials
share qualitative features with such quadratic potential, see \emph{e.g.}
\cite{singh2006nonsingular}, therefore we expect that obtained results
are generic for inflationary scenarios in Loop Quantum Cosmology.

{\bf Acknowledgments:} we thank both referees for their remarks, which allowed to improve presentation of our results.


\begin{thebibliography}{10}

\bibitem{guth1981inflationary}
A.~H. Guth,
\newblock Physical Review D {\bf 23}, 347 (1981).

\bibitem{linde1982anew}
A.~Linde,
\newblock Physics Letters B {\bf 108}, 389 (1982).

\bibitem{albrecht1982cosmology}
A.~Albrecht and P.~J. Steinhardt,
\newblock Physical Review Letters {\bf 48}, 1220 (1982).

\bibitem{linde1983chaotic}
A.~Linde,
\newblock Physics Letters B {\bf 129}, 177 (1983).

\bibitem{mukhanov1992theoryof2}
V.~F. Mukhanov, H.~A. Feldman, and R.~H. Brandenberger,
\newblock Physics Reports {\bf 215}, 203  (1992).

\bibitem{belinsky1985inflationary}
V.~A. Belinskii, L.~P. Grishchuk, I.~M. Khalatnikov, and Y.~B. Zeldovich,
\newblock Sov. Phys. JETP {\bf 62}, 195 (1985).

\bibitem{1985PhLB..155..232B}
V.~A. {Belinsky}, L.~P. {Grishchuk}, I.~M. {Khalatnikov}, and Y.~B.
  {Zeldovich},
\newblock Physics Letters B {\bf 155}, 232 (1985).

\bibitem{gibbons1987anatural}
G.~Gibbons, S.~Hawking, and J.~Stewart,
\newblock Nuclear Physics B {\bf 281}, 736 (1987).

\bibitem{schiffrin2012measure}
J.~S. Schiffrin and R.~M. Wald,
\newblock Physical Review D {\bf 86} (2012).

\bibitem{hawking1988howprobable}
S.~Hawking and D.~N. Page,
\newblock Nuclear Physics B {\bf 298}, 789 (1988).

\bibitem{carroll2010unitary}
S.~M. Carroll and H.~Tam,
\newblock arXiv:1007.1417 [astro-ph, physics:gr-qc, physics:hep-th]  (2010),
\newblock arXiv: 1007.1417.

\bibitem{gibbons2008measure}
G.~W. Gibbons and N.~Turok,
\newblock Physical Review D {\bf 77} (2008).

\bibitem{brandenberger2017initial}
R.~Brandenberger,
\newblock International Journal of Modern Physics D {\bf 26}, 1740002 (2017).

\bibitem{remmen2013attractor}
G.~N. Remmen and S.~M. Carroll,
\newblock Physical Review D {\bf 88} (2013).

\bibitem{corichi2014inflationary}
A.~Corichi and D.~Sloan,
\newblock Classical and Quantum Gravity {\bf 31}, 062001 (2014).

\bibitem{1987ZhETF..93..784B}
V.~A. {Belinskii} and I.~M. {Khalatnikov},
\newblock Sov. Phys. JETP {\bf 66}, 441 (1987).

\bibitem{borde2003inflationary}
A.~Borde, A.~H. Guth, and A.~Vilenkin,
\newblock Physical Review Letters {\bf 90} (2003).

\bibitem{page1984afractal}
D.~N. Page,
\newblock Classical and Quantum Gravity {\bf 1}, 417 (1984).

\bibitem{thiemann2003lectures}
T.~Thiemann,
\newblock Lectures on {Loop} {Quantum} {Gravity},
\newblock in {\em Quantum {Gravity}}, edited by R.~Beig et~al., volume 631,
  pages 41--135, Springer Berlin Heidelberg, Berlin, Heidelberg, 2003.

\bibitem{ashtekar2004background}
A.~Ashtekar and J.~Lewandowski,
\newblock Classical and Quantum Gravity {\bf 21}, R53 (2004).

\bibitem{rovelli2004quantum}
C.~Rovelli,
\newblock {\em Quantum {Gravity}},
\newblock Cambridge University Press, Cambridge, 2004.

\bibitem{bojowald2008loopquantum}
M.~Bojowald,
\newblock Living Reviews in Relativity {\bf 11} (2008).

\bibitem{ashtekar2006quantum}
A.~Ashtekar, T.~Pawlowski, and P.~Singh,
\newblock Physical Review D {\bf 74} (2006).

\bibitem{ashtekar2006quantum2}
A.~Ashtekar, T.~Pawlowski, and P.~Singh,
\newblock Physical Review D {\bf 73} (2006).

\bibitem{singh2006nonsingular}
P.~Singh, K.~Vandersloot, and G.~V. Vereshchagin,
\newblock Physical Review D {\bf 74} (2006).

\bibitem{2009CQGra..26l5005S}
P.~{Singh},
\newblock Classical and Quantum Gravity {\bf 26}, 125005 (2009).

\bibitem{ashtekar2010loopquantum}
A.~Ashtekar and D.~Sloan,
\newblock Physics Letters B {\bf 694}, 108 (2010).

\bibitem{corichi2011measure}
A.~Corichi and A.~Karami,
\newblock Physical Review D {\bf 83} (2011).

\bibitem{1989NYASA.571..249P}
R.~{Penrose},
\newblock Annals of the New York Academy of Sciences {\bf 571}, 249 (1989).

\bibitem{linsefors2013duration}
L.~{Linsefors} and A.~{Barrau},
\newblock \prd {\bf 87}, 123509 (2013).

\bibitem{bolliet_clarifications_2017}
B.~{Bolliet}, A.~{Barrau}, K.~{Martineau}, and F.~{Moulin},
\newblock Classical and Quantum Gravity {\bf 34}, 145003 (2017).

\bibitem{2011CQGra..28u3001A}
A.~{Ashtekar} and P.~{Singh},
\newblock Classical and Quantum Gravity {\bf 28}, 213001 (2011).

\bibitem{domagala2004blackhole}
M.~Domagala and J.~Lewandowski,
\newblock Class. Quantum Grav. {\bf 21}, 5233 (2004).

\bibitem{meissner2004blackhole}
K.~A. Meissner,
\newblock Classical and Quantum Gravity {\bf 21}, 5245 (2004).

\bibitem{singh2005semiclassical}
P.~Singh and K.~Vandersloot,
\newblock Physical Review D {\bf 72} (2005).

\bibitem{taveras2008corrections}
V.~Taveras,
\newblock Physical Review D {\bf 78} (2008).

\bibitem{banerjee2005discreteness}
K.~Banerjee and G.~Date,
\newblock Class. Quantum Grav. {\bf 22}, 2017 (2005).

\bibitem{vereshchagin2003flatcosmological}
G.~V. Vereshchagin,
\newblock International Journal of Modern Physics D {\bf 12}, 1487 (2003).

\bibitem{bojowald2001absence}
M.~Bojowald,
\newblock Physical Review Letters {\bf 86}, 5227 (2001).

\bibitem{remmen2014howmany}
G.~N. {Remmen} and S.~M. {Carroll},
\newblock \prd {\bf 90}, 063517 (2014).

\end{thebibliography}

\end{document}